\documentclass[usenatbib]{mn2e}

\usepackage{epsfig,latexsym,graphicx,lscape}

\defcitealias{ATel2107}{Markwardt et al. 2009a}
\defcitealias{ATel2120}{Markwardt et al. 2009b}
\defcitealias{ATel2108}{Markwardt \& Swank 2009}

\begin{document}

\title[XTE J1652$-$453 in the hard-intermediate state]{A strong and broad Fe line in the \textit{XMM-Newton} spectrum of the new X-ray transient and black-hole candidate XTE J1652$-$453}
\author[B. Hiemstra et al.]
       {Beike Hiemstra$^{1}$\thanks{E-mail: hiemstra@astro.rug.nl}, 
        Mariano M\'{e}ndez$^{1}$, Chris Done$^{2}$, Mar\'{i}a D\'{i}az Trigo$^{3}$,
\newauthor Diego Altamirano$^{4}$, and Piergiorgio Casella$^{5}$\\
        $^{1}$Kapteyn Astronomical Institute, University of Groningen, P.O. Box 800, 9700 AV Groningen, The Netherlands\\
        $^{2}$Department of Physics, University of Durham, South Road, Durham, DH1 3LE, UK\\
        $^{3}$\textit{XMM-Newton} Science Operation Centre, ESAC, P.O. Box 78, 28691 Villanueva de la Ca\~nada, Madrid, Spain\\	
        $^{4}$Astronomical Institute Anton Pannekoek, University of Amsterdam, Science Park 904, 1098 XH Amsterdam, The Netherlands\\
        $^{5}$School of Physics and Astronomy, University of Southampton, Southampton, Hampshire, SO17 1BJ, UK}
\date{Submitted on April 23 2010 (v1); Accepted on September 7 2010 (this version. v3) }

\pagerange{\pageref{firstpage}--\pageref{lastpage}}
\pubyear{2010}

\maketitle
\label{firstpage}

\begin{abstract}
We observed the new X-ray transient and black-hole candidate XTE J1652$-$453 
simultaneously with \textit{XMM-Newton} and the \textit{Rossi X-ray Timing Explorer} 
(\textit{RXTE}). The observation was done during the decay of the 2009 outburst, when 
XTE J1652$-$453 was in the hard-intermediate state. The spectrum shows a strong and 
broad Fe emission line with an equivalent width of $\sim 450$ eV. The profile is 
consistent with that of a line being produced by reflection off the accretion disc, 
broadened by relativistic effects close to the black hole. The best-fitting inner 
radius of the accretion disc is $\sim 4$ gravitational radii. Assuming that the 
accretion disc is truncated at the radius of the innermost stable circular orbit, the 
black hole in XTE J1652$-$453 has a spin parameter of $\sim 0.5$. The power spectrum 
of the \textit{RXTE} observation has an additional variability component above 50 Hz, 
which is typical for the hard-intermediate state. No coherent quasi-periodic oscillations 
at low frequency are apparent in the power spectrum, which may imply that we view the 
system at a rather low inclination angle.
\end{abstract}

\section{Introduction}\label{sec:intro}
Most binaries with a stellar-mass black-hole primary are X-ray transients in which 
matter is accreted from the companion star onto the black hole (BH). These systems 
are characterised by a sudden and rapid rise in intensity (reaching a maximum in 
about one to several weeks) followed by a gradual decline. The so-called outbursts 
typically last several months; eventually the source becomes too weak to be detected 
with current X-ray instruments \citep[for typical light curves, see Fig. 4--9 in][]{Remillard06}. 
The large excursions in the intensity seen in X-ray transients are likely due to 
changes in the accretion rate onto the BH, triggered by disc instabilities due to 
hydrogen ionisation \citep[see Section 2.2.1 in][]{Done07,Lasota01}.

During an outburst, not only the intensity, but also the X-ray spectrum and X-ray variability 
changes \citep[and references therein]{Mendez97,McClintock2006,Belloni10}. In the bright phase 
of an outburst, at X-ray luminosities $L_{\rm X} \sim10^{37}$--$10^{39}$ erg s$^{-1}$, the X-ray spectrum 
is dominated by thermal emission below $\sim 5$ keV originating from an optically thick and 
geometrically thin accretion disc, with a typical temperature of $\sim 1$ keV, that extends 
down to the marginally stable orbit \citep{Shakura73}. This phase is generally referred to as 
the soft or high/soft state, and is further characterised by a low level of 
variability, with typical fractional rms values of a few per cent. When the luminosity 
decreases to about 1--4 per cent of the Eddington luminosity \citep[][]{Maccarone03}, the source 
makes a transition to the so-called hard or low/hard state, in which the X-ray spectrum is dominated by hard 
power-law like emission with a typical power-law photon index of 1.5--1.7. The high-energy emission 
is produced by inverse Compton scattering of photons by a population of hot electrons in a region 
referred to as the corona \citep{Thorne75}. The hard state is further characterised by a high 
level of variability, with rms amplitudes of $> 20$ per cent, with strong quasi-periodic oscillations 
(QPO) at low frequencies \citep[$\sim 0.1$--10 Hz; e.g.,][]{Homan05}.

X-ray spectra of BH transients sometimes show broad Fe emission lines at around 6--7 keV 
\citep[for a review see][]{Miller07}. These lines, from fluorescent-K$\alpha$ emission of 
neutral or ionised Fe, are thought to be produced through irradiation of the disc by the 
source of hard photons. The Fe line is broadened by relativistic effects and becomes 
asymmetric \citep{Fabian89,Laor91}, and its profile provides information on the disc 
properties, such as the inclination angle, the emissivity, and the inner radius. The latter 
can be used to measure the BH spin \citep[e.g.,][]{Reynolds08}, assuming that the disc extends 
down to the innermost stable circular orbit (ISCO) in the soft states.

In this paper we report on the timing and spectral analysis of a contemporaneous \textit{XMM-Newton} 
and \textit{RXTE} observation of the new X-ray transient XTE J1652$-$453. This is the first (and so 
far the only) time that this source has been observed with an instrument with a good spectral 
resolution to study the Fe line which was previously detected in \textit{RXTE} Proportianal Counter 
Array (PCA) data \citepalias{ATel2107}. Furthermore, the broadband coverage of the simultaneous 
\textit{XMM-Newton} and \textit{RXTE} data allows to properly model the underlying continuum spectrum. 
In Section~\ref{sec:obs&red} we describe the outburst evolution from \textit{RXTE} observations, 
we give details about the \textit{XMM-Newton} and \textit{RXTE} observations, and we describe the 
steps that we followed to generate the power and energy spectra. We continue with the data analysis 
in Section~\ref{sec:data analysis} where we test different line and continuum models, and where we 
consider several broadening mechanisms. In this section we also present our results from the spectral 
and timing analysis separately. In Section~\ref{sec:discussion} we discuss our results in the context 
of the spectral state, disc geometry, and inferred black hole spin. In the appendix we discuss how the 
choice of background spectrum and abundances influences the continuum and line parameters inferred 
from the spectrum.
\subsection{The 2009 outburst of XTE J1652$-$453}\label{sec:2009outburst}
XTE J1652$-$453 was discovered in the \textit{RXTE}/PCA Galactic bulge scan on June 28 2009 \citepalias{ATel2107}.
Follow-up observations with the \textit{Swift} X-ray Telescope (XRT) and 
\textit{RXTE}/PCA measured a spectrum consistent with a stellar-mass black hole in the canonical 
high/soft state \citepalias{ATel2108,ATel2120}, with the spectrum below 10 keV dominated by emission 
from an accretion disc with a temperature of $\sim 0.6$ keV. The \textit{RXTE}/PCA spectrum also 
showed evidence of a relatively strong emission line at 6.4 keV, most likely originating from an 
Fe-K$\alpha$ line. However, the line could not be studied in detail because of the limited spectral 
resolution of \textit{RXTE}.

A near infrared (NIR) imaging observation of the field of XTE J1652$-$453 on July 11 2009, 
provided a possible NIR counterpart of XTE J1652$-$453 \citep{ATel2125}. However, from NIR 
observations on July 15 and August 28 2009, the suggested NIR counterpart showed no significant 
variability, suggesting that the previous detection might have been an unrelated interloper star 
\citep{ATel2190}. Radio observations with the Australia Telescope Compact Array on July 6, 7, 
and 13 2009, showed emission that may be indicative of the decay of an optically thin synchrotron 
event associated with the activation of XTE J1652$-$453 \citep{ATel2135}.

On August 22 2009, we triggered an \textit{XMM-Newton} observation of XTE J1652$-$453, simultaneously 
with \textit{RXTE}, to study the Fe emission line and the disc properties in the decline of the outburst. 
In this paper we present the results of these observations.

Around September 10 2009, the Burst Alert Telescope (BAT) on-board \textit{Swift} showed an increase (by 
a factor of 1.5--2) in the hard X-ray count rate (15--50 keV) of XTE J1652$-$453, indicating a 
transition to the hard state \citep{ATel2219}. On September 14 2009, the \textit{RXTE} spectrum was still 
consistent with a disc blackbody ($kT \sim 0.6$ keV), a power law (index of $\sim 1.8$), and a line 
feature at $\sim 6$ keV. Later \textit{RXTE} observations did not require a disc component, and no line 
feature was apparent \citep{ATel2219}.
\section{Observations \& Data reduction}\label{sec:obs&red}
\subsection{Hardness and intensity as function of time}\label{sec:intensity}
Since its discovery, XTE J1652$-$453 was observed with \textit{RXTE} every $\sim 3$--4 days. There are
in total 55 \textit{RXTE} observations of XTE J1652$-$453, each of them lasting from $\sim 1$ to several ks.

We used the 16-s time-resolution \textrm{Standard 2} data from the PCA \citep[for instrument information see][]{Jahoda06} 
to calculate the hard colour and intensity of all 55 observations. For each of the five PCA units 
(PCUs), and for each time interval of 16 s, we calculated the intensity, defined as the count rate 
in the 2--20 keV band, and the hard colour, defined as the count rate in the 16--20 keV range divided 
by the count rate in the 2--6 keV range.

To correct for differences in effective area between the PCUs and for time-dependent gain changes of
the detectors, we used the method described in \citet{Straaten03}. As we did for XTE J1652$-$453, 
for each PCU we calculated the hard colour and intensity of the Crab nebula, which are assumed to 
be constant, and averaged the 16-s Crab colours and intensities per day. For each PCU, we divided 
the 16-s hard colours and intensities of XTE J1652$-$453 by the corresponding averaged Crab values 
that were closest in time. Finally, we averaged the 16-s colours and intensities over all PCUs per 
observation.

We further used the 15--50 keV fluxes of XTE J1652$-$453 measured by \textit{Swift}/BAT during the outburst. 
These data are provided by the \textit{Swift}/BAT team and are publicly available on the \textit{Swift}/BAT 
transient monitor page~\footnote{\texttt{http://swift.gsfc.nasa.gov/docs/swift/results/transients/}}.
\subsection{Energy spectrum of XTE J1652$-$453}
XTE J1652$-$453 was simultaneously observed with \textit{XMM-Newton} and \textit{RXTE} on August 22 2009; both 
observations started at the same time (03:39 UTC). The \textit{XMM-Newton} observation (obsID 0610000701) 
lasted for $\sim 42$ ks, while the \textit{RXTE} observation (obsID 94432-01-04-00) lasted for $\sim 14$ 
ks. To achieve a broadband spectral coverage, we extracted the energy spectra of XTE J1652$-$453 
from the \textit{XMM-Newton} data as well as from the \textit{RXTE} observation. We describe the data reduction 
of the individual instruments in the following sections.
\subsubsection{\textit{XMM-Newton}}\label{sec:XMM-Newton}
For the data reduction of the \textit{XMM-Newton} Observational Data Files (ODF) we used the Science 
Analysis Software (SAS) version 9.0, applying the latest calibration files. We generated 
concatenated and calibrated event lists for the different instruments, using the standard
SAS tasks. The European Photon Imaging Camera (EPIC) PN and MOS2 were operated in timing 
mode, while MOS1 was operated in imaging mode. The Reflection Grating Spectrometers (RGS) 
were set in normal spectroscopy mode.

For the EPIC data we applied the standard filtering criteria, as described in the science 
threads provided by the \textit{XMM-Newton} team, and we additionally filtered out the flaring 
particle background, leading to a net exposure of $\sim 35$ ks. Except for the flaring
background, the light curve was rather constant during the whole observation. Since MOS1 
was operated in imaging mode, it was heavily affected by photon pile-up. After excising the 
pile-up affected central region of the source, the statistics of the MOS1 spectrum were 
strongly reduced, and we therefore did not include this spectrum in the rest of our analysis. 
Using the task \textsc{epatplot}, we evaluated the pile-up fraction in MOS2 and PN data, and 
we found that no photon pile-up was apparent for MOS2 and PN. To further check whether the 
spectrum was affected by pile-up \citep[see e.g.,][]{Yamada09,DoneTrigo09}, we extracted 
the source spectra excluding the central three columns of the point spread function (PSF). 
There are no significant differences between the spectrum that includes the core of the PSF 
and the one that excludes it. 

Following the standard procedure, we extracted source and background spectra for MOS2 and PN 
data, and we created redistribution matrices and ancillary files using the SAS tools \textsc{rmfgen}
and \textsc{arfgen}, respectively. The MOS2 source spectrum was extracted using the RAWX columns in 
288--328, and for single events only (\textsc{pattern} 0). We extracted the MOS2 background from 
a source-free circular region on the outer CCDs, $\sim 15$ arcmin away from the source, using 
\textsc{pattern} 0, 1, and 3.

The standard correction for charge-transfer inefficiency (CTI) does not work properly for PN 
operated in fast mode (burst or timing) and bright sources, which can lead to a gain shift 
and may affect the energy spectrum \citep{Sala06}. Therefore, before extracting the PN spectra, 
we used the new SAS task \textsc{epfast} to correct for possible CTI effects. (More information 
about CTI correction can be found at: \texttt{http://xmm.esac.esa.int/external/xmm\_calibration/.}) 
We extracted the PN source spectrum using RAWX in [30:46]; for the PN background we initially 
used the RAWX columns in [5:21]. However, the background spectrum appeared to be contaminated 
with source photons (see Fig.~\ref{fig:src-vs-bkg}). We tried different background extraction 
regions, but since the wings of the PSF extends beyond the CCD boundaries, the background 
spectra taken from the same CCD were always contaminated with source photons. Using this 
contaminated background would lead to an over-subtraction of the source spectrum, and may 
affect the results of the spectral analysis. Since during our observation the source count
rate was significantly higher than that of the background (source $\sim 100$ cts s$^{-1}$; no 
contaminated background $< 0.5$ cts s$^{-1}$), we decided not to subtract any background 
\citep[see also][]{DoneTrigo09}. However, we investigated whether the best-fitting parameters 
are affected by applying different background corrections. We therefore performed a number of 
spectral fits using different background models. Details about these fits can be found in the 
Appendix.

The EPIC spectra obtained following the standard procedures oversample the intrinsic resolution 
of the detectors by a factor of $\sim 10$. We therefore rebinned the EPIC spectra with the tool 
\textsc{pharbn}~\footnote{\texttt{http://virgo.bitp.kiev.ua/docs/xmm\_sas/Pawel/reduction/pharbn}}, 
to have 3 energy channels per resolution element, and at least 15 counts per channel. We did not 
apply any systematic error to the EPIC spectra.	

We reduced the RGS data and produced all necessary files following the standard procedure. We 
included the RGS spectra at low energies (below 2 keV), only using the first order spectra of 
RGS1 and RGS2. Unless mentioned differently, we rebinned the energy channels above $\sim 0.9$ 
keV by a factor of 10, and the channels below that energy such that there are at least 15 counts 
per bin. No systematic error was applied to the RGS data.
\subsubsection{\textit{RXTE}}
For the \textit{RXTE} data reduction we used the \textsc{heasoft} tools version 6.7. We followed the 
standard procedure suggested by the \textit{RXTE} team~\footnote{\texttt{http://heasarc.gsfc.nasa.gov/docs/xte/xtegof.html}}. 
After standard screening and filtering we obtained a net exposure of $\sim 13$ and $\sim 4$ ks 
for PCA and High Energy X-ray Timing Experiment (HEXTE), respectively. Using the tool \textsc{saextrct}, 
we extracted the PCA spectrum from \textrm{Standard-2} data, using only events from PCU2, being this 
the best-calibrated detector. We extracted the HEXTE spectra using cluster-B events only. The PCA 
background was estimated and the HEXTE background was measured with the tools \textsc{pcabackest} 
and \textsc{hxtback}, respectively. We generated PCA and HEXTE instrument response files, using 
\textsc{pcarsp} and \textsc{hxtrsp}, respectively. Within the used energy range the energy channels 
of the \textit{RXTE} data contained more than 15 counts per bin, such that $\chi^{2}$ statistics 
can be applied, and no channel rebinning was required. No systematic errors were added to the HEXTE 
data, while we used $0.6$ per cent systematic errors for the PCA data.
\subsection{Power spectrum of XTE J1652$-$453}\label{sec:PDS}
We used the \textit{RXTE}/PCA data to calculate power density spectra of all available observations of 
XTE J1652$-$453. For this we used the Event-mode data with 125-$\mu$s time resolution over the 
nominal 2--60 keV range, without any energy selection, using data from all available PCUs. For 
all 55 \textit{RXTE} pointed observations, we first produced light curves at a 1-s time resolution to 
identify and exclude drop outs in the data. We then divided each observation into segments of 
continuous 128 s, and we calculated the Fourier transform of each segment preserving the available 
time resolution. Therefore, each power spectrum extends from 1/128 to 4096 Hz, with a frequency 
resolution of 1/128 Hz. Finally, we averaged all the 128-s power spectra within an observation
to produce an averaged power density spectrum per observation. 

For each observation we subtracted the dead-time modified Poisson level following the prescription 
in \citet{Zhang95}. Using the corresponding source and background count rates, we re-normalised 
the averaged power spectra of each observation to rms units \citep{Vanderklis89}. We then 
calculated the integrated fractional rms amplitude for each observation in the frequency range 
1/128--100 Hz.

In addition, for the \textit{RXTE} observation (ObsID 94432-01-04-00) that was coincident with the \textit{XMM-Newton}
observation, we fit the power spectrum with a function consisting of the sum of three Lorentzians 
\citep[e.g.,][]{Belloni02}. The results of this fit are described in Section~\ref{sec:timing}.
\section{Data analysis and results}\label{sec:data analysis}
\subsection{Outburst evolution}\label{sec:evolution}
In Fig.~\ref{fig:lightcurve} we show the intensity and hard colour as a function of time for the 
outburst of XTE J1652$-$453. We also show the BAT flux and the 1/128--100 Hz fractional rms 
amplitude. The first detection of this source was on MJD 55010 \citepalias{ATel2107}. The PCA pointed 
observations started $\sim 4$ days later, when XTE J1652$-$453 had an intensity of about 80 mCrab. 
In the following 2 days the intensity increased, reaching a maximum of about 140 mCrab at MJD 
$\sim 55016$. From that point onwards, the intensity of the source decayed more or less exponentially 
until MJD $\sim 55065$. After that day the intensity remained approximately constant ($\sim 20$ mCrab) 
for about 55 days, showing a dip at around MJD 55080. Starting on MJD 55120, the intensity started to 
decrease again, and $\sim 25$ days later it dropped down below the detection limit of \textit{RXTE}. 
In total, the outburst lasted approximately 145 days.

From the beginning of the outburst the hard colour increased slowly, roughly linearly with time 
until MJD $\sim 55065$, and at that point it started to increase more rapidly. The faster increase 
occurred at the same time as the intensity stopped its exponential decay (see above). After this, 
the hard colour continued increasing, showing a small dip at MJD 55080, and it further increased 
by a factor of $\sim 3$ within $\sim 6$ days. The time of the sudden increase in the hard colour 
coincides with the dip in the intensity curve. Starting on MJD 55087, and for about 13 days, the 
hard colour remained approximately constant. After MJD 55120 the hard colour started to increase 
again, at the time the intensity started to decrease again, until XTE J1652$-$453 became too weak 
to be detected with \textit{RXTE}. The sudden increase in the hard colour at around MJD 55080 also coincides 
with a slight increase of the BAT flux light curve, as shown in the third panel of Fig.~\ref{fig:lightcurve}.

The moment of our \textit{XMM-Newton} observation is marked by a filled black circle in Fig.~\ref{fig:lightcurve}. 
Our observation samples the outburst at the moment that the intensity stabilised at about $\sim 20$ mCrab, 
a day before the first increase in the hard colour.

In the bottom panel of Fig.~\ref{fig:lightcurve} we show the 1/128--100 Hz fractional rms amplitude 
as a function of time during the outburst of XTE J1652$-$453. In the first few observations the 
rms amplitude is low, at around 2--4 per cent. As the outburst progresses, and the source intensity
decreases, the rms amplitude increases more or less linearly with time, up to $\sim 30$ per cent on 
MJD 55094, then it decreases slightly to $\sim$15--20 per cent for about a week, and then increases 
again up to 30--40 per cent at the end of the outburst. At the time of the simultaneous \textit{XMM-Newton}
observation, on MJD 55065, the 1/128--100 Hz fractional rms amplitude of XTE J1652$-$453 was 
somewhat intermediate, $13.3 \pm 1.2$ per cent. 
\setcounter{figure}{0}
\begin{figure}
\begin{center}
\resizebox{\columnwidth}{!}{\rotatebox{-90}{\includegraphics[]{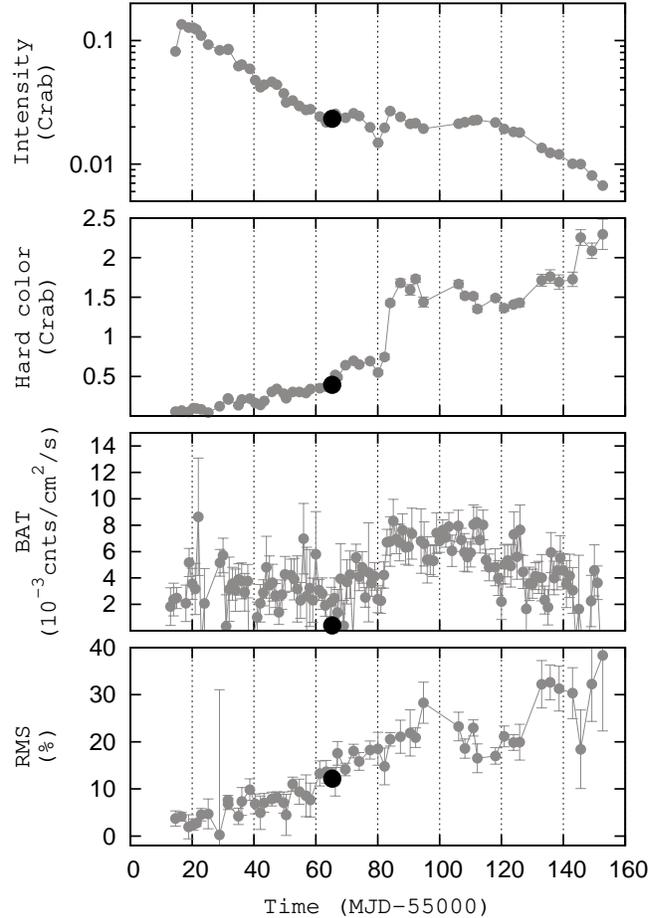}}}
\caption{From top to bottom: Intensity (2--20 keV), Hard colour (16--20 keV/ 2--6 keV), BAT flux 
(15--50 keV), and fractional rms amplitude (1/128--100 Hz, full PCA band), respectively. Intensities, 
hard colours and fractional rms amplitudes are the average per day of each pointed PCA observation. 
The BAT measurements were taken from the BAT transient monitoring. The black filled circle marks the 
moment of the \textit{XMM-Newton} observation.}
\label{fig:lightcurve}
\end{center}
\end{figure}
\subsection{Spectral state at the moment of the \textit{XMM-Newton} observation}
From Fig.~\ref{fig:lightcurve} it is apparent that at the beginning of the outburst XTE J1652$-$453 
was in the high/soft state, with high intensity, low hard colour, and low 1/128--100 Hz fractional 
rms amplitude \citep[e.g.,][]{Mendez97,Homan05}. Around MJD 55080, there is a clear transition to the 
low/hard state, with a sharp increase of the hard colour, while both the BAT flux and the 1/128--100 Hz 
fractional rms amplitude increase slightly, but significantly. Transitions from the high/soft state to 
the intermediate state \citep[soft- or hard-intermediate;][]{Homan05,Belloni10} are generally smooth 
transitions, and without multi-wavelength observations in combination with detailed timing analysis it 
is difficult to say when this transition exactly happened.

In Fig.~\ref{fig:HID} we show the hardness intensity diagram (HID) for the outburst of XTE J1652$-$453. 
For comparison we also show the HID of the black-hole candidate GX 339$-$4 of the 2002/2003 outburst. 
Both diagrams show the intensities and hard colours in Crab units, both calculated in the same manner 
as described in Section~\ref{sec:intensity}. In the HID of GX 339$-$4 we marked the different spectral 
states identified by \citet{Belloni05} and \citet{Homan05}. Suggested by Fig.~\ref{fig:HID} together with 
the 1/128--100 Hz integrated rms amplitude of $\sim 15$ per cent, which is too high for the soft or the
soft-intermediate state, we conclude that at the moment of the \textit{XMM-Newton} observation, 
XTE J1652$-$453 was in the hard-intermediate state.
\setcounter{figure}{1}
\begin{figure}
\begin{center}
\hspace{-0.68cm}
\resizebox{1.07\columnwidth}{!}{\rotatebox{0}{\includegraphics[clip]{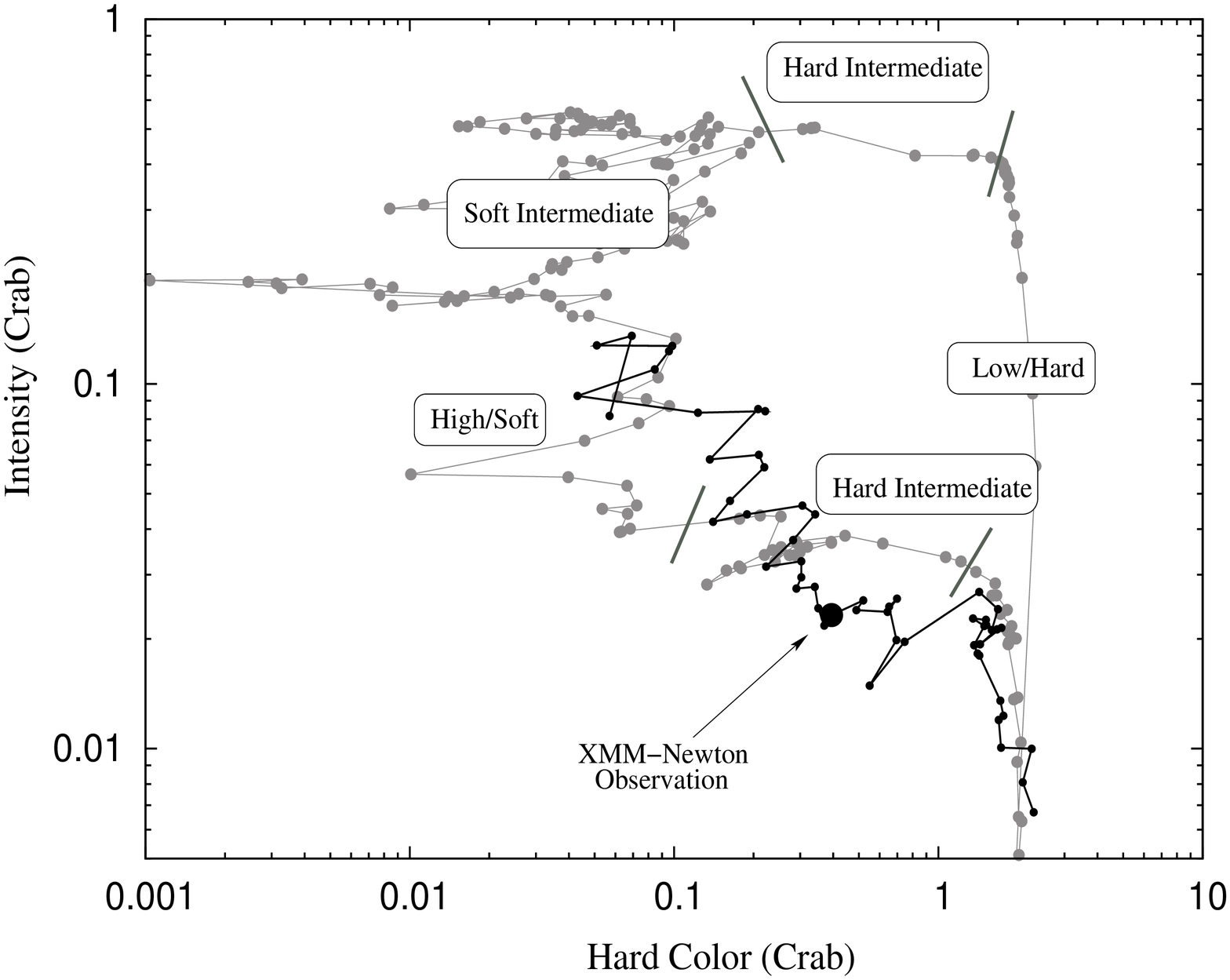}}}
\caption{Hardness intensity diagram of the 2009 outburst of XTE J1652$-$453 (black), and of the 2002/2003 
outburst of GX 339$-$4 (grey). Intensities and hard colours are given in Crab units (for more details see 
Section~\ref{sec:intensity}). For the HID of GX 339$-$4 we marked the different states identified as in 
\citet{Belloni05} and \citet{Homan05}. We also marked the moment of the \textit{XMM-Newton} observation.}
\label{fig:HID}
\end{center}
\end{figure}
\subsection{Spectroscopy}\label{sec:spectroscopy}
For the spectral analysis of XTE J1652$-$453 we used \textsc{xspec} version 12.5.1 \citep{Arnaud96}. 
Since \textit{XMM-Newton} and \textit{RXTE} started their observations at the same time, together with the 
rather constant light curve during the whole \textit{XMM-Newton} observation, we fit the \textit{XMM-Newton} 
and \textit{RXTE} spectra simultaneously. We initially used a continuum model including a power law 
(\textsc{powerlaw} in \textsc{xspec}) and a disc blackbody \citep[\textsc{diskbb};][]{Mitsuda84}.
We included a multiplicative factor in our fit to account for differences in the absolute flux 
calibration between the different instruments, and to account for interstellar absorption we used 
the photoelectric absorption component \textsc{phabs}, assuming cross sections of \citet{Verner96} 
and abundances of \citet{Wilms00}. (In the appendix we show how the choice of cross sections and 
abundances may affect the parameters.)

A strong and broad emission line at around 6--7 keV was apparent in the spectra, most likely due 
to Fe fluorescence lines (of neutral or ionised Fe), and we initially fit the line with a Gaussian. 
However, a Gaussian line did not provide a good fit ($\chi_{\nu}^{2} \simeq 2.6$); the Gaussian fit 
a large part of the line between 6--7 keV, but it was not able to fit the red wing extending down 
to 4--5 keV. Guided by the asymmetric profile, we fit the line with a relativistic line profile 
\citep[e.g., for a Kerr-metric: \textsc{Laor};][]{Laor91}. We tried different line and continuum 
components to test whether the line parameters depend on the choice of line and/or continuum model. 
Details and results of this can be found in Section~\ref{sec:line&continuum}.

When fitting the \textit{XMM-Newton}/\textit{RXTE} spectra, several features appeared in the fitting 
residuals. Some of them are due to calibration issues of \textit{XMM-Newton} data in timing mode, 
some of them are caused by cross-calibration issues between the different instruments, and others 
might be real. We discuss these features individually in the following Section.

Unless stated differently, all uncertainties given in this paper are at a 90 per cent confidence 
level ($\Delta \chi^{2} = 2.71$), and upper limits are given at a 95 per cent confidence level. 
The equivalent widths (EW) of emission and absorption features are both quoted with positive values.

\subsubsection{Instrumental/Cross-calibration issues}\label{sec:calibration}
Here we list all features and issues that arose during our spectral analysis which are most 
likely related with uncertainties in the (cross) calibration of the different instruments.\\\\
\textit{Excess at $\sim 1$ keV:}

When we fit the PN spectrum in the 0.7--11 keV range with a line and continuum model, \textsc{phabs(powerlaw + diskbb + Laor)}, 
we found a strong and broad emission feature at $\sim 1$ keV, extending up to $\sim 1.3$ keV 
(see Fig.~\ref{fig:1keVexcess}). The $\sim 1$ keV excess was also apparent in the MOS2 data, 
although the profile of the excess was different. Since both detectors showed this excess we 
cannot discard that it may be real, and we considered the possibility that it might be due to 
blurred Fe-L or Ne-X emission lines \citep[see][]{Fabian09}. We tried to fit this excess with 
either a Gaussian component or a relativistic line profile, but none of these models gave a 
good fit. No excess at $\sim 1$ keV was apparent in the RGS data, although the lower effective 
area does not allow us to discard it completely.
\setcounter{figure}{2}
\begin{figure}
\begin{center}
\resizebox{\columnwidth}{!}{\rotatebox{-90}{\includegraphics[clip]{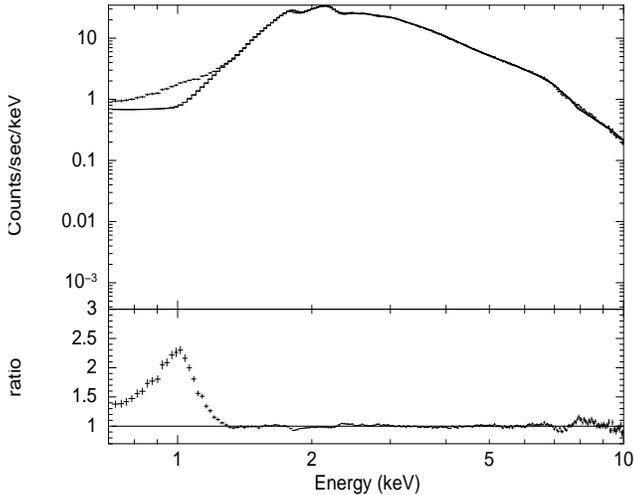}}}
\caption{An excess at $\sim 1$ keV is apparent in the PN spectrum of XTE J1652$-$453. We show
the data/model ratio when we fit the data with a \textsc{phabs(powerlaw + diskbb + Laor)} model. The 
spectrum is not corrected for background, and instrumental residuals are still present.}
\label{fig:1keVexcess}
\end{center}
\end{figure}

Similar excesses at $\sim 1$ keV have been seen in other sources observed with PN operated in 
timing mode \citep[e.g.,][]{Martocchia06,Sala08,Boirin05}. The excess appears in that part of 
the spectrum where the flux drops abruptly due to interstellar absorption, and the EW of the 
excess appears to increase as a function of this absorption \citep[see e.g.,][]{DoneTrigo09}. 
Given this, a calibration issue related with the redistribution matrix of the timing-mode data 
cannot be excluded at the moment. Due to the uncertain origin of this excess (a thorough 
investigation is beyond the scope of this paper), we ignored the EPIC data below 1.3 keV and 
included the RGS data to cover the spectrum below that energy.\\\\
\textit{Energy-dependent discrepancy between PCA and PN:}

Fitting the PCA data in the 3--20 keV range shows an energy-dependent deviation between the PN 
and PCA spectra below $\sim 7$ keV. (With respect to the PN data, the PCA data show an excess 
that increases with decreasing energy.) Fitting the PN and PCA spectra individually (in the 
1.3--11 keV and 3--20 keV ranges, respectively), there is a discrepancy in the power-law photon 
index, which is most likely due to a mismatch in cross-calibration between the PN and PCA 
instruments. To avoid the effects of this apparent energy-dependent discrepancy between PN 
and PCA data, in the rest of our analysis we only used PCA data in the 7--20 keV range.\\\\
\textit{Absorption features:}

The fit residuals of the \textsc{phabs(powerlaw + diskbb + Laor)} model to the EPIC data show two 
absorption features at around $1.8$ and $2.1$ keV, consistent with the Si-K (1.84 keV) and Au-M 
(2.21 keV) edges, which are most likely of instrumental origin. To reduce the impact of these 
lines, we could fit each of them with a Gaussian absorption component at $1.85 \pm 0.01$ and 
$2.13 \pm 0.02$ keV, and widths of $55 \pm 5$ eV and $110 \pm 10$ eV, respectively, although 
some residuals between $\sim 1.8$--2.2 keV still remained, probably due to problems with the 
gain and the CTI correction. In order to keep the model as simple as possible, instead of fitting 
these two instrumental features with absorption components, we ignored all energy channels between 
1.75--2.25 keV so that these instrumental edges will not affect the results of the analysis.
However, we caution that the edges, if caused by residual problems of the CTI correction, indicate 
that a slight energy shift may appear in the whole energy band (see \textit{XMM-Newton} Calibration 
Documentation XMM-SOC-CAL-TN-0083, on \texttt{http://xmm.vilspa.esa.es/docs/documents/}).

The RGS data showed an absorption feature at $\sim 1.4$ keV. In case we fit this absorption 
line with a \textsc{gabs} component, the best-fitting line energy is $1.42 \pm 0.06$ keV, line width
$\sigma=70 \pm 10$ eV, and EW of $\sim 35$ and $\sim 62$ eV for RGS1 and RGS2, respectively. 
The width and strength of this absorption feature suggest that it is most likely due to a 
mismatch in the spectrum curvature of the PN and RGS data. In the rest of the analysis we fit 
this absorption line with a \textsc{gabs} component.

Further, the EPIC fit residuals above 6 keV showed additional features in the PN spectrum at 
high energy (see Fig.~\ref{fig:absorption}). The slight increase at $\sim 8.3$ keV (and the 
consequent dip at higher energies) is probably due to the emission lines in the detector 
background (see Fig.~\ref{fig:src-vs-bkg}). Since the absorption feature at $\sim 8.7$ keV is 
only apparent in the PN data, and not in the MOS2 or PCA data, it is most likely an effect of 
not correcting for the background emission lines in the PN detector. To improve the fit statistics, 
we fit the absorption line at $8.78 \pm 0.07$ keV, with width $\sigma=70\pm20$ eV and EW of $37 \pm 10$ eV. 
\setcounter{figure}{3}
\begin{figure}
\begin{center}
\resizebox{\columnwidth}{!}{\rotatebox{-90}{\includegraphics[clip]{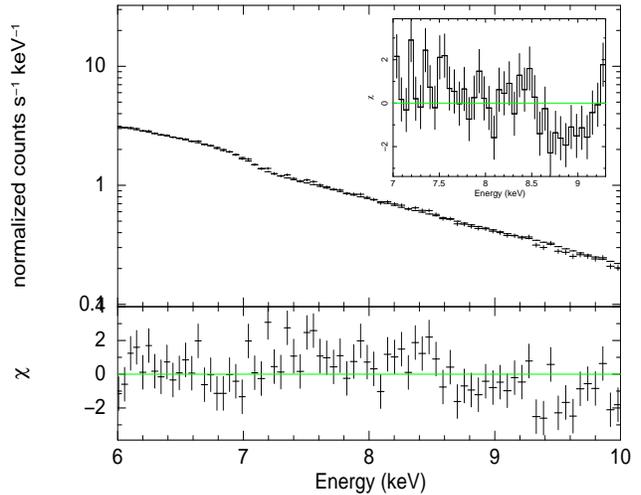}}}
\caption{Zoom in of the PN spectrum of XTE J1652$-$453. The absorption feature at $\sim 9.2$ keV 
is apparent in the lower panel, where we show the residuals when fitting the data with the 
\textsc{phabs(powerlaw + diskbb + Laor)} model. In the sub-panel we show a zoom in of the residuals 
to show the apparent absorption features at $\sim 8.5$--9 keV.}
\label{fig:absorption}
\end{center}
\end{figure}

In contrast to all above mentioned absorption features, the dip at $\sim 9.2$ keV could be real 
since both PN and PCA spectra show this feature. Since the highly ionised Fe line should be 
accompanied by an associated edge, we fit this feature with an \textsc{edge} component, and found 
a best-fitting edge depth of $\tau_{max}=0.20\pm0.02$ at an energy of $9.05^{+0.12}_{-0.07}$ keV, 
midway between the He and H-like edges of Fe at $\sim 8.83$ and $\sim 9.28$ keV, respectively. For 
PCA data only, the best-fitting $\tau_{max}=0.19\pm0.04$ at an energy of $9.14\pm0.17$ keV, 
consistent with the edge of H-like Fe. 
\subsubsection{Line and Continuum models}\label{sec:line&continuum}
For testing different line and continuum models, we simultaneously fit PN in the 1.3--11 keV 
range (excluding the 1.75--2.25 keV range), RGS in the 0.4--2 keV range, PCA in the 7--20 keV 
range, and HEXTE in the 20--150 keV range. From the available \textit{XMM-Newton}/EPIC data we 
only included PN data since this is the best calibrated instrument in timing mode, but for a 
consistency check we also fit the MOS2 data separately (see Section~\ref{sec:MOS2}).\\\\
\textit{Different relativistic line profiles:}

The broad and skewed Fe emission line profile that is apparent in the spectra of XTE J1652$-$453 
(see Fig.~\ref{fig:Feline}) is consistent with what is expected from relativistic effects, and 
we therefore fit it with the commonly used line profiles for a non-rotating \citep[\textsc{diskline};][]{Fabian89} 
and a maximally rotating black hole (\textsc{Laor}).
Both line components have similar parameters: line energy, $E_{L}$, which depends on the ionisation 
stage of Fe, inner disc radius, $R_{\rm in}$, outer disc radius, $R_{\rm out}$, disc inclination, 
$\theta$, and emissivity index, $\beta$, which is the power-law dependence of the disc emissivity 
that scales as $R^{-\beta}$. The normalisation of the line, $N_{\rm L}$, gives the number of 
photons cm$^{-2}$ s$^{-1}$ in the line. We let all parameters free, except $R_{\rm out}$ which we 
fixed to 400 $GM/c^{2}$ (with $G$ and $c$ the usual physical constants, and $M$ the mass of the 
black hole). We further constrained the line energy to the range between 6.4--6.97 keV, since this 
is the expected range for neutral to hydrogen-like Fe-K$\alpha$ lines.

The fit with the \textsc{diskline} component did not provide a good fit with $\chi^2_{\nu} \sim 1.5$ 
for 362 d.o.f., with strong residuals below 6 keV, and the inner radius pegged at 6 $GM/c^{2}$. Since 
the \textsc{diskline} appears not to be the right model to describe the line, we do not further mention 
any results for this model. 
The \textsc{Laor} profile however, gives an acceptable fit with $\chi^2_{\nu} \sim 1.2$ for 362 d.o.f. 
In Table~\ref{tab:line-models} we show the best-fitting parameters for the model with \textsc{Laor}, 
with a \textsc{diskbb} and \textsc{powerlaw} as our continuum. In the rest of the paper we refer 
to the \textsc{phabs(diskbb + powerlaw + Laor)} model as Model 1. The continuum parameters are the 
power-law photon index, $\Gamma$, power-law normalisation, $N_{\rm \Gamma}$ (in photons keV$^{-1}$ 
cm$^{-2}$ s$^{-1}$ at 1 keV), equivalent hydrogen column density, $N_{\rm H}$, temperature at the 
inner disc radius, $kT_{\rm in}$, and the disc blackbody normalisation, 
$N_{\rm dbb}=[(R_{\rm in}$/km)/($D$/10 kpc)$]^{2} \cos \theta$, where $R_{\rm in}$ is the apparent 
inner disc radius, $D$ the distance to the source, and $\theta$ the angle of the disc. 

The broadband spectrum of XTE J1652$-$453 and the fit residuals for Model 1 are shown in the left 
panel of Fig.~\ref{fig:Feline}, with in the inset the residuals in the Fe line region when fitting 
the underlying continuum spectrum with Model 1 (except the line component) and excluding the Fe 
line band, i.e., the 4--7 keV range. In the right panel of Fig.~\ref{fig:Feline} we show the unfolded 
spectrum and the model components of Model 1.
\setcounter{figure}{4}
\begin{figure*}
\begin{center}
\hspace{-0.3cm}
\resizebox{\columnwidth}{!}{\rotatebox{-90}{\includegraphics[clip]{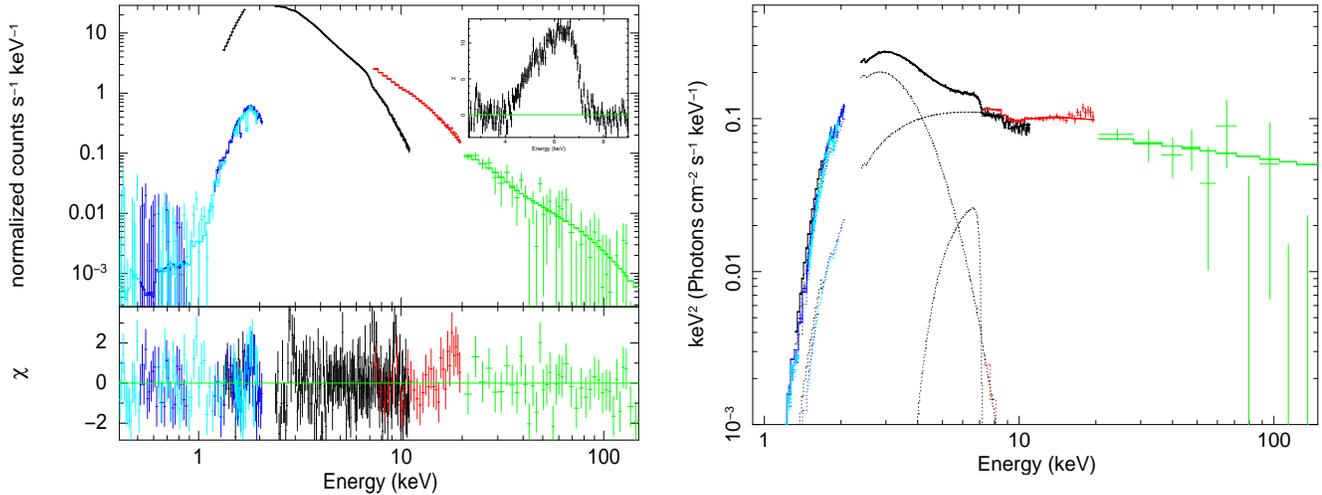}}}
\hspace{0.4cm}
\resizebox{\columnwidth}{!}{\rotatebox{-90}{\includegraphics[clip]{EEUF-model1-HEXTEbin.ps}}}
\caption{\textit{Left:} \textit{XMM-Newton}/\textit{RXTE} spectra of XTE J1652$-$453: RGS (0.4--2 keV, 
blue), PN (1.3--11 keV, black), PCA (7--20 keV, red), and HEXTE (20--150 keV, green) with, in the bottom 
panel, the best-fitting residuals of Model 1. In the sub-panel in the upper-right corner we show the 
residuals in the Fe line region when fitting the underlying continuum spectrum with Model 1 (without 
\textsc{Laor}) and excluding the Fe line band, i.e., the 4--7 keV range. \textit{Right:} Unfolded 
\textit{XMM-Newton}/\textit{RXTE} spectra of XTE J1652$-$453 and the model components (Model 1). For 
plotting purposes we binned the HEXTE data by a factor or 4.}
\label{fig:Feline}
\end{center}
\end{figure*}

Since the \textsc{Laor} profile assumes a maximally spinning black hole, we also tried a relativistic 
line profile that allows to fit for the black hole spin \citep[\textsc{kyrline}, up-dated version, 
private communication with M. Dov\v{c}iak, but see also,][]{Dovciak04}. Besides the black hole angular 
momentum, $a/M$, the \textsc{kyrline} parameters are $\theta$, $R_{\rm in}$, $R_{\rm out}$, $E_{\rm L}$, 
and $N_{\rm L}$, which are the same parameters (and same units) as for \textsc{Laor}. We fixed 
$R_{\rm out}$ to 400 $GM/c^{2}$ and left all other parameters free. Further, within \textsc{kyrline} 
it is possible to set the directionality law for the emission (e.g., isotropic or Limb 
darkening/brightening), and to have a broken power-law radial emissivity profile. We assumed a 
disc with a single emissivity index, where $\beta$ was free to vary. We further assumed a transfer 
function including Laor's limb darkening, and we only included emission from outside the marginally 
stable orbit. We fit \textsc{kyrline} using the same continuum as before, which is our Model 2: 
\textsc{phabs(diskbb + powerlaw + kyrline)}. The best-fitting parameters for this model are also given
in Table~\ref{tab:line-models}.

From Table~\ref{tab:line-models} (columns 1 and 2) and Fig~\ref{fig:Feline} we can see that the 
spectrum of XTE J1652$-$453 is well described by three components. The spectrum shows a significant
contribution of thermal emission coming from a disc with a temperature of $\sim 0.6$ keV, where the 
disc fraction is $\sim 75$ per cent. The spectrum has significant hard emission up to at least 40 keV, 
where the hard tail is described by a power law with a photon index of $\sim 2.2$. We further find 
that that the column density of interstellar matter in the direction of XTE J1652$-$453 is 
$\sim 7 \times 10^{22}$ cm$^{-2}$, which absorbs most of the X-rays coming from the disc. From 
Fig.~\ref{fig:Feline} it is clear that there is a strong asymmetric Fe emission line with a red 
wing extending down to about 4 keV. From fits with the relativistic line profiles we find a line 
that is more than $15\sigma$ significant (based on the $1\sigma$ error of the line flux), and a 
best-fitting line energy of 6.97 keV. Despite the fact that the \textsc{kyrline} model is more 
complex (having more parameters to fit for) and has a better grid resolution (more steps over 
which the line is integrated) than the \textsc{Laor} component, the inferred line parameters for 
both profiles are still consistent. From both relativistic line profiles we find that the disc is 
observed at a low inclination, $< 20^{\circ}$, and has an inner disc radius of $\sim 3.5$ $GM/c^{2}$. 
From the \textsc{kyrline} profile we find that the Fe emission line is indicative of a fast 
spinning black hole, although it is also consistent with a moderately spinning black hole, 
with a spin parameter of $\sim 0.45$.\\\\
\textit{Different continuum models:}

As shown in previous work \citep[see e.g.,][]{Hiemstra09}, the choice of continuum model can 
affect the parameter values of the line and other continuum components, affecting any physical 
interpretation. To investigate the effect of the continuum on the inferred line profile, we 
tested different continuum components. Given that the best-fitting line parameters are 
consistent between the fits with a \textsc{Laor} or a \textsc{kyrline} component, we used the 
\textsc{Laor} profile to model the Fe emission line since it gives a slightly better fit, and 
it is computational much faster than \textsc{kyrline}.\\

\textit{I: \textsc{cutoffpl}}\\
We first tried to fit the hard emission with a \textsc{cutoffpl} component to see whether there is
an exponential roll-off at high energy. However, we did not find a roll-off in the hard power-law 
tail within the energy coverage of our data. Therefore, using a \textsc{cutoffpl} is effectively 
the same as using a \textsc{powerlaw} component. \\

\textit{II: \textsc{CompTT}}\\
Next (Model 3), we replaced the \textsc{powerlaw} component in Model 1 with a component that 
describes the hard emission as thermal Comptonisation of soft photons in a hot plasma 
\citep[\textsc{CompTT};][]{Titarchuk94}, where we assumed a disc geometry for the Comptonising 
plasma. The parameters of \textsc{CompTT} are the input soft photon (Wien) temperature $T_{0}$, 
the plasma temperature $kT$, the plasma optical depth $\tau$, and the normalisation $N_{\rm Comp}$. 
The best-fitting results for Model 3 are given in Table~\ref{tab:line-models}. 

We find that the spectrum of XTE J1652$-$453 is well described by a thermal component and a 
Comptonising component, where the Comptonising plasma is hot ($kT \sim 170$ keV) and optically thin 
($\tau \ll 1$). Because of the low-energy cutoff at the seed photon energy, the unabsorbed 0.5--10 keV 
flux of the \textsc{CompTT} component is $\sim 2$--3 times lower than the flux of the \textsc{powerlaw} 
component in Models 1 and 2. Due to this cutoff, the hard power-law tail does not contribute to
very low energies and causes the slightly lower value for $N_{\rm H}$, and thus a slightly higher
contribution of the disk component. The other parameters are consistent with the values obtained 
in Models 1 and 2.\\

\textit{III: \textsc{CompPS}}\\
Subsequently, we tried a component that describes a hybrid thermal/non-thermal Comptonising plasma 
\citep[\textsc{CompPS};][]{Poutanen96}. This model component computes the Comptonisation spectra 
taking into account the geometry and optical depth of the corona, the spectral distribution of soft 
seed photons, and the parameters of the electron distribution. The resulting spectrum includes the 
incident thermal emission, the Comptonized emission, and the Compton reflection. For the source of 
soft seed photons (the incident thermal emission) we assumed a multicolor disk blackbody, and we 
therefore did not include a separate \textsc{diskbb} component to the model. Since we only wanted 
to test how a hybrid plasma affects the line and continuum parameters, we did not include the Compton 
reflection in \textsc{CompPS}, but we still added a \textsc{Laor} component to fit the Fe line. For a 
spherical geometry of the corona, the best-fitting results for this model (Model 4) are given in 
Table~\ref{tab:line-models}. The free parameters are the electron temperature of the corona, $kT_{e}$, 
seed photon temperature, $kT_{dbb}$, optical depth of the corona, $\tau$, and normalization, $N_{CPS}$, 
which has the same units as the \textsc{diskbb} normalization. We further fixed the power-law index of 
the energy distribution of the non-thermal electrons to 2, we set the minimum and maximum lorentz factors 
of the non-thermal electrons to 2 and 3000, respectively, and set the abundances to solar.

We find that Model 4 fits the data well with $\chi^{2}_{\nu} \sim 1.2$ for 362 d.o.f. The hybrid 
Comptonising plasma has a best-fitting temperature of $\sim 75$ keV and is optically thin ($\tau \sim 0.15$). 
The hybrid plasma is cooler than the fully thermal plasma in Model 3, while the optical depth is a 
factor 2 larger. We further find that the parameters of the disk and line are hardly affected by the 
choice of the Comptonising component (non-thermal, Model 1, thermal, Model 3, or hybrid, Model 4), with 
values being consistent within errors. However, the equivalent width is $\sim 10$--20 per cent lower in 
Model 4 than in the other 2 models.\\

\textit{IV: \textsc{reflionx}}\\
Since the line energy of the Fe emission line suggests a high ionisation stage of Fe, we tried an 
ionised reflection model \citep[\textsc{reflionx};][]{Ross05}. This model describes the reflection 
off an ionised and optically-thick atmosphere of constant density (such as the surface of an 
accretion disc) that is illuminated by radiation with a power-law spectrum. Since \textsc{reflionx} 
only provides the reflected emission, we included a \textsc{powerlaw} component to account for the 
direct emission. We coupled the photon index of the illuminating power-law component in \textsc{reflionx} 
to the photon index of the \textsc{powerlaw}. \textsc{reflionx} includes ionisation states and 
transitions from the most important ions, like O III-VIII, Fe VI-XXVI, and several others. To 
account for relativistic effects in the disc, we used a relativistic convolution component 
\citep[\textsc{Kerrconv};][]{Brenneman06}. All together, Model 5 was \textsc{phabs(diskbb + powerlaw + Kerrconv$\ast$reflionx)} 
model. For this continuum model the absorption component at $\sim 8.7$ keV was not significant 
anymore and we therefore took it out. The absorption component at $\sim 1.4$ keV and the edge 
at $\sim 9.2$ keV were still included. The free parameters in the \textsc{reflionx} model are 
the ionisation parameter, $\xi$ (where $\xi=4\pi F/n$, with $F$ the total illuminating flux 
and $n$ the hydrogen number density), and the normalisation, $N_{\rm ref}$. We initially also 
let the Fe abundance to be free, finding a best-fitting value of 4 times solar. This is unlikely 
for a low-mass X-ray binary, and we therefore fixed the Fe abundance to solar. For \textsc{Kerrconv}, 
we assumed a single emissivity index, which was free to vary. The black hole spin and inclination 
angle of the disc were also free parameters. We further fixed the inner radius of the disc to be 
equal to the radius of the marginally stable orbit, $R_{\rm ms}$, and the outer radius to 400 
$R_{\rm ms}$. The best-fitting parameters for Model 5 are also given in Table~\ref{tab:line-models}.

We find that Model 5 fits the overall continuum and Fe emission line quite well, except that a clear 
absorption feature remained in the residuals at $\sim 7.1$ keV, causing a slightly higher $\chi^{2}_{\nu}$ 
compared with Models 1--3. Although this absorption feature is not obviously apparent in the raw data, 
we checked whether it can be either a K-edge of neutral Fe (7.11 keV), a blue-shifted absorption line of 
highly ionised Fe, or a mismatch in the model used to describe the data. We tried to fit the residuals at 
$\sim 7.1$ keV with an \textsc{edge} component, but since the feature is narrow, an edge does not provide 
a good fit. When we fit the residuals with a \textsc{gabs} component, the best-fitting absorption line 
has a central energy of 7.2 keV and a width of 0.1 keV. For the case of a blue-shifted Fe-XXVI Ly-$\alpha$ 
line, the blue shift corresponds to a velocity of $\sim 9500$ km s$^{-1}$, which is significantly higher 
than the blue shift of $\sim 300$--1600 km s$^{-1}$ observed in GRO J1655$-$40 \citep{MillerNature06}. For 
other Fe lines the blue shift would be even larger. Since \textsc{reflionx} self-consistently incorporates
Compton scattering, the Fe feature in this model has an extended blue wing up to $\sim 8$ keV, whereas 
the observed Fe emission line in the spectrum of XTE J1652$-$453 shows a sharp drop at $\sim 7$ keV. We 
therefore conclude that the negative residuals at $\sim 7.1$ keV are likely caused by a deficiency in 
the model \citep[see][]{Done06}.

From Model 5 we find a high ionisation parameter, $\xi \sim 2900$ erg cm s$^{-1}$, indicating that 
the emission line is dominated by highly ionised Fe, which is consistent with the line energy 
of 6.97 keV that we inferred for the relativistic line profiles in Models 1--3. We also find a 
moderately spinning black hole with a spin parameters of $\sim 0.5$, consistent with the lower 
confidence value inferred from Model 2. Further, the best-fitting inclination is consistent 
with that inferred from \textsc{Laor}, while the best-fitting emissivity index is slightly lower 
than we found from Models 1--3. The \textsc{powerlaw} component in Model 5 only represents the 
direct emission, and therefore its normalisation is $\sim 2$ times lower than for the other three 
models. All other parameters of Model 5 are consistent with those inferred for the other models. 
The reflection fraction for Model 5, as estimated from the (0.5--150 keV) flux ratio between the 
reflected and incident emission, is $\sim 90$ per cent.
\setcounter{table}{0}
\begin{table*}
\begin{minipage}{160mm}
\begin{center}{
\scriptsize \caption{Different line and continuum models fit to \textit{XMM-Newton}/\textit{RXTE} data 
of XTE J1652$-$453}\label{tab:line-models}
\begin{tabular}{clccccc}

\multicolumn{1}{c}{Model component} & 
\multicolumn{1}{c}{Parameter} & 
\multicolumn{1}{c}{\hrulefill~Model 1~\hrulefill} &  
\multicolumn{1}{c}{\hrulefill~Model 2~\hrulefill} &  
\multicolumn{1}{c}{\hrulefill~Model 3~\hrulefill} & 
\multicolumn{1}{c}{\hrulefill~Model 4~\hrulefill} & 
\multicolumn{1}{c}{\hrulefill~Model 5~\hrulefill} \\ \hline\hline

\textsc{phabs}    & $N_{\rm H}$ ($10^{22}$)   & 6.73$\pm$0.01          & 6.75$\pm$0.02          & 6.66$\pm$0.01          & 6.71$\pm$0.02    & 6.73$\pm$0.01 \\\hline

\textsc{diskbb}   & $kT_{\rm in}$ (keV)       & 0.59$\pm$0.01          & 0.59$\pm$0.01          & 0.60$\pm$0.01          & --               & 0.59$\pm$0.01 \\
                  & $N_{\rm dbb}$             & 1015$^{+36}_{-11}$     & 1051$\pm$40            & 1090$^{+16}_{-36}$     & --               & 998$\pm$10    \\
                  & Flux ($10^{-9}$)          & 2.09$\pm$0.07          & 2.12$\pm$0.08          & 2.37$^{+0.20}_{-0.08}$ & --               & 2.08$\pm$0.02 \\\hline

\textsc{powerlaw} & $\Gamma$                  & 2.23$\pm$0.03          & 2.22$\pm$0.03          & --                     & --               & 2.16$\pm$0.02 \\
                  & $N_{\rm \Gamma}$          & 0.18$\pm$0.01          & 0.18$\pm$0.01          & --                     & --               & 0.09$\pm$0.01 \\
                  & Flux ($10^{-9}$)          & 0.74$\pm$0.06          & 0.73$\pm$0.05          & --                     & --               & 0.39$\pm$0.04 \\\hline

\textsc{CompTT}   & $T_{0}$ (keV)             & --                     & --                     & 0.88$^{+0.12}_{-0.03}$ & --               & -- \\
                  & $kT$ (keV)                & --                     & --                     & 170$^{+24}_{-51}$      & --               & -- \\
                  & $\tau$                    & --                     & --                     & 0.06$\pm$0.01          & --               & -- \\
                  & $N_{\rm Comp}$ ($10^{-4}$)& --                     & --                     & 3.25$^{+9.31}_{-1.12}$ & --               & -- \\ 
                  & Flux ($10^{-9}$)          & --                     & --                     & 0.27$^{+0.77}_{-0.09}$ & --               & -- \\\hline

\textsc{CompPS}   & $kT_{e}$ (keV)            & --                     & --                     & --                     & 75.3$\pm$1.3           & -- \\
                  & $kT_{dbb}$ (keV)          & --                     & --                     & --                     & 0.58$\pm$0.01          & -- \\
                  & $\tau$                    & --                     & --                     & --                     & 0.15$\pm$0.01          & -- \\
                  & $N_{CPS}$                 & --                     & --                     & --                     & 1314$\pm$37            & -- \\
                  & Flux ($10^{-9}$)          & --                     & --                     & --                     & 2.72$^{+0.12}_{-0.03}$ & -- \\\hline

\textsc{Laor}     & $E_{\rm L}$ (keV)         & 6.97$^{+0}_{-0.02}$    & --                     & 6.97$^{+0}_{-0.06}$    & 6.97$^{+0}_{-0.02}$    & -- \\
                  & $\beta$                   & 3.4$\pm$0.1            & --                     & 3.4$\pm$0.1            & 3.3$\pm$0.1            & -- \\
                  & $R_{\rm in}$ ($GM/c^{2}$) & 3.65$^{+0.31}_{-0.02}$ & --                     & 3.64$^{+0.27}_{-0.03}$ & 4.01$^{+0.18}_{-0.10}$ & -- \\ 
                  & $\theta$ (deg)            & 9.4$^{+4.5}_{-3.8}$    & --                     & 9.4$\pm$3.9            & 5.3$^{+3.3}_{-5.3}$    & -- \\ 
                  & $N_{\rm L}$ ($10^{-3}$)   & 1.6$\pm$0.1            & --                     & 1.6$\pm$0.1            & 1.3$\pm$0.1            & -- \\
                  & EW (eV)                   & 455$^{+17}_{-26}$      & --                     & 451$^{+21}_{-28}$      & 366$^{+21}_{-15}$      & -- \\\hline

\textsc{kyrline}  & $a/M$ ($GM/c$)            & --                     & 0.96$^{+0.01}_{-0.51}$ & --                     & --                   & -- \\
                  & $E_{\rm L}$ (keV)         & --                     & 6.97$^{+0}_{-0.04}$    & --                     & --                   & -- \\
                  & $\beta$                   & --                     & 3.5$\pm$0.1            & --                     & --                   & -- \\
                  & $R_{\rm in}$ ($GM/c^{2}$) & --                     & 3.39$^{+0.31}_{-0.19}$ & --                     & --                   & -- \\
                  & $\theta$ (deg)            & --                     & 16.6$\pm$3.0           & --                     & --                   & -- \\
                  & $N_{\rm L}$ ($10^{-3}$)   & --                     & 1.8$\pm$0.2            & --                     & --                   & -- \\
                  & EW (eV)                   & --                     & 470$^{+28}_{-41}$      & --                     & --                   & -- \\\hline

\textsc{reflionx} & $\xi$ (erg cm s$^{-1}$)   & --                     & --                     & --                     & --              & 2872$^{+807}_{-26}$\\
                  & $N_{\rm ref}$ ($10^{-7}$) & --                     & --                     & --                     & --              & 4.9$^{+0.04}_{-0.23}$\\\hline

\textsc{Kerrconv} & $a/M$                     & --                     & --                     & --                     & --              & 0.45$\pm$0.02 \\
                  & $\beta$                   & --                     & --                     & --                     & --              & 2.7$\pm$0.1 \\
                  & $\theta$ (deg)            & --                     & --                     & --                     & --              & 8.8$\pm$0.1 \\\hline

        & $\chi^{2}_{\nu}$ ($\chi^{2}/\nu$)   & 1.22 (422.2/362)       & 1.25 (451.1/361)       & 1.21 (437.2/360)       & 1.23 (445.2/362)  & 1.26 (460.2/365)\\
        & Total Flux ($10^{-9}$)              & 2.84$\pm$0.04          & 2.87$\pm$0.03          & 2.65$^{+0.76}_{-0.09}$ & 2.73$\pm$0.02     & 2.86$\pm$0.11 \\\hline\hline

\multicolumn{7}{c}{~~~NOTES.--~Results are based on spectral fits to RGS/PN/PCA/HEXTE data in the 0.4--150 keV range. Columns 1--2 give the results of}\\ 
\multicolumn{7}{c}{different relativistic line models using the continuum model \textsc{phabs(diskbb + powerlaw)} with line profiles \textsc{Laor} (Model 1) and ~\textsc{kyrline} (Mo-}\\ 
\multicolumn{7}{c}{del 2). Columns 3--5 give the results of different continuum models. Models 3 and 4, respectively, include a thermal Comptonising plasma}\\
\multicolumn{7}{c}{(\textsc{CompTT}) and a hybrid plasma (\textsc{CompPS}) to model the hard emission, and the \textsc{Laor} component is used to fit the Fe line. For Model 4 the}\\
\multicolumn{7}{c}{Comptonising spectrum self-consistently includes the incident thermal emission, so we did not include a separate \textsc{diskbb} component as we}\\
\multicolumn{7}{c}{did for Model 3. ~Model 5 is the continuum model \textsc{phabs(diskbb + powerlaw + Kerrconv$\ast$reflionx)}, where \textsc{reflionx} describes the reflected}\\
\multicolumn{7}{c}{emission, \textsc{powerlaw} models the direct emission, and \textsc{Kerrconv} smears for relativistic effects. For details about parameters, see text. Cross-}\\
\multicolumn{7}{c}{calibration constants for the different instruments with respect to PN are: 1.09 (PCA), 0.84 (RGS1 \& HEXTE), and 0.79 (RGS2). Fluxes}\\
\multicolumn{7}{l}{~~are unabsorbed and given for the 0.5--10 keV range in units of erg cm$^{-2}$ s$^{-1}$.}

\end{tabular}
\normalsize}
\end{center}
\end{minipage}
\end{table*}
\subsubsection{Consistency check: MOS2 data}\label{sec:MOS2}
For consistency, we verified our results obtained from fits to the RGS/PN/\textit{RXTE} data (in this 
section simply referred to as PN data) with fits to the RGS/MOS2/\textit{RXTE} data (and hence referred 
to as MOS2 data). For comparison we only fit Model 1 to the MOS2 data, assuming the same abundances, 
cross sections, and energy ranges that we used to fit the PN data (see Section~\ref{sec:spectroscopy}). 
We ignored the MOS2 channels between 1.75--2.25 keV for similar reasons as for PN, added a Gaussian 
absorption component to fit the $\sim$1.4 keV feature seen in the RGS spectra, and fit the $\sim 9.2$ 
keV feature in MOS2 with an \textsc{edge}.

Model 1 (including all above absorption features) fit the MOS2 data well with $\chi^{2}_{\nu} = 1.25$ 
(458.0/366). The unabsorbed 0.5--10 keV flux is $2.3 \times 10^{-9}$ erg cm$^{-2}$ s$^{-1}$, 
and the best-fitting parameters for the continuum components are $N_{\rm H}=(6.77\pm0.05) \times 10^{22}$ cm$^{-2}$, 
$kT_{\rm in}=0.60\pm0.01$ keV, $N_{\rm dbb}=859^{+59}_{-47}$, $\Gamma=2.17\pm0.05$, and 
$N_{\Gamma}=0.13\pm0.02$ photons keV$^{-1}$ cm$^{-2}$ s$^{-1}$ at 1 keV. For the Fe line the 
best-fitting values are $E_{\rm L}=6.97^{+0}_{-0.51}$ keV, $R_{\rm in}=2.8\pm0.5$ $GM/c^{2}$, 
$\beta=3.7\pm0.4$, $\theta=25^{+18}_{-6}$ degrees, and $N_{\rm L}=1.9 \pm 0.5 \times10^{-3}$ photons 
cm$^{-2}$ s$^{-1}$ in the line. The Fe emission line has an EW of $514^{+100}_{-75}$ eV consistent 
with that obtained from PN data. The edge at $\sim 9.2$ and the absorption feature at $\sim 1.4$ 
keV have parameters which are consistent with the values obtained from the PN data.

We find that most of the continuum and line parameters obtained from the MOS2 data are consistent 
(within errors) with the values obtained from the PN data. The only parameters that were not consistent 
are the \textsc{diskbb} normalisation and the inclination. However, the inclination inferred from the 
MOS2 data is consistent (within errors) with the inclination deduced from the \textsc{kyrline} profile 
that we fit to the PN data. The \textsc{diskbb} normalisation obtained from fits to the MOS2 data is 
lower than that for PN, which is most likely due to the fact that we subtracted the background spectrum 
in the case of the MOS2 data, but we did not for the PN data.
\subsection{Other possible broadening mechanisms}
By using the \textsc{Laor} and \textsc{kyrline} components in our Models 1--4 we are 
in fact assuming that the line is \emph{only} broadened by relativistic effects. However, 
one should consider the possibility that other mechanisms contribute to the total line 
width. Besides broadening due to thermal Comptonisation by the disk, as in \textsc{reflionx} 
(see Section~\ref{sec:line&continuum}), an alternative mechanism could be Compton broadening due 
to electron scattering in the hot plasma above the accretion disc \citep{Czerny91}. The width of 
a Compton scattered line depends on the temperature and the optical depth of the Comptonising plasma. 
For a single Compton scattering $\sigma_{\rm E}/E = [(2kT_{\rm e})/(m_{\rm e} c^{2})]^{1/2}$, 
with $\sigma_{\rm E}$ the line width, $E$ the central line energy, $kT_{\rm e}$ the electron 
temperature of the plasma, and $m_{\rm e}$ and $c$ the usual physical constants \citep{Pozdnyakov83}. 
From fits including a \textsc{CompTT} component to describe the high-energy emission (Model 3), 
we find $\tau \simeq 0.06$ and $kT_{\rm e} \simeq 170$ keV (see Table~\ref{tab:line-models}), which 
means that the Comptonising plasma would only intercept a fraction of 0.06 of the line photons, 
and scatter these into a broader line of width $\sim 5.5$ keV. A fraction of 0.94 of the line 
is unaffected, and comes out at its intrinsic width. Thus the line would have a predominantly 
narrow core, but with 6 per cent of the photons scattered to such extent that it would be 
difficult (or even impossible) to distinguish from the continuum. However, as a matter of 
fact the problem here is that in all probability the Comptonisation is \textit{not} thermal but 
instead is non-thermal. Therefore the obtained optical depth and temperature are not real. 
As we found from the fits of a hybrid Comptonisation model (Model 4), the plasma can also be 
cooler ($kT_{\rm e} \simeq 75$ keV), but we still found a low optical depth ($\tau \sim 0.15$). 
In this case $\sim 15$ per cent of the line photons will be scattered into a broader line of width 
$\sim 3.5$ keV, which is still too broad and too weak to be distinguished from the continuum.

\citet{Kallman89} showed that the apparent widths of Fe lines may be due to a blend of 
narrow line components arising from different ionisation states, with each of these narrow 
components in turn affected by Compton broadening. The combination of blending and Compton 
broadening produces a line profile with an equivalent width, line width and line energy that 
depend on several factors, besides the ionisation parameter of the emitting gas. In this 
scenario, the total line width is not only set by relativistic effects, but also by the 
combined effects of gas motions (rotation), Compton scattering, and blending of emission
of several ionisation states, where the centroid line energy is a function of the degree of 
ionisation of Fe, and hence the ionisation parameter of the emitting gas \citep{Kallman89}.
To estimate the effect of line blending, we fit the Fe line with a multi-component line 
model. Assuming the continuum of Model 1, we added three \textsc{Laor} components with 
the line energies fixed at 6.4, 6.7 and 6.97 keV, for neutral, He-like, and H-like Fe, 
respectively. We fixed the outer radius of the disc to 400 $GM/c^{2}$, and let the rest 
of the parameters free to vary. Except for the line normalisations, we coupled the line 
parameters of each of the three \textsc{Laor} profiles. We find that the multi-line model 
fits the observed Fe emission line well, $\chi^{2}_{\nu}=1.17$ for 361 d.o.f. The best-fitting 
parameters are $R_{\rm in}=2.20^{+0.10}_{-0.03}$ $GM/c^{2}$, $\beta=4.0 \pm 0.1$, 
$\theta=40.7 \pm 1.5$ degrees, and $N_{\rm L}= 2.0 \pm 0.2 \times 10^{-3}$, and 
$0.6 \pm 0.28 \times 10 ^{-3}$ photons cm$^{-2}$ s$^{-1}$ for the 6.4 and 6.97 keV lines, 
respectively. The 6.7 keV line was not required by the fit, with UL to the EW of $\sim 51$ 
eV. The equivalent widths of the neutral and H-like Fe were $389^{+20}_{-39}$ and $127 \pm 41$ 
eV, respectively. The best-fitting continuum parameters were consistent with the values 
reported for a single-line model (Model 1, Table~\ref{tab:line-models}). From this we 
initially concluded that line blending can partially explain the observed line profile, 
although relativistic effects are still important. The resulting disc inclination was 
significantly higher for the multi-line model than for the single-line model. In contrast 
to the single \textsc{Laor} profile, the multi-line model was dominated by the neutral-Fe 
line, with an equivalent width $\sim 3$ times higher than the one of the H-like Fe line. 
Based on the high ionisation parameter, $\xi \sim 2900$ erg cm s$^{-1}$, obtained from the 
fits with \textsc{reflionx} (see Table~\ref{tab:line-models}) we do not expect a significant 
fraction of neutral Fe in the disc, which casts some doubt on this approach. Moreover, given 
the observed high disc temperature we do not expect much neutral Fe, irrespective of photo-ionisation.

A third broadening mechanism that we considered explains the red-skewed emission line by down-scattering 
of line photons in an optically thick and partly ionised wind from the BH \citep[see][and references therein]{Titarchuk09}. 
In this model the effect of photo-absorption of the line photons at higher energies (above 7--8 keV, 
depending on the ionisation stage of the wind) suppresses the blue wing of the Fe emission line, and 
therefore leads only to a redshifted profile, where the redshift of the line photons is to first order 
a function of $v/c$, with $v$ the outflow velocity. Contrary to the optically thick Comptonising wind 
assumed in this model, we find an optically thin corona for the case of XTE J1652$-$453, which makes this 
outflow model unlikely to explain the observed broad and red-skewed Fe line.
\subsection{Timing analysis}\label{sec:timing}
\setcounter{figure}{5}
\begin{figure}
\begin{center}
\resizebox{\columnwidth}{!}{\rotatebox{0}{\includegraphics[clip]{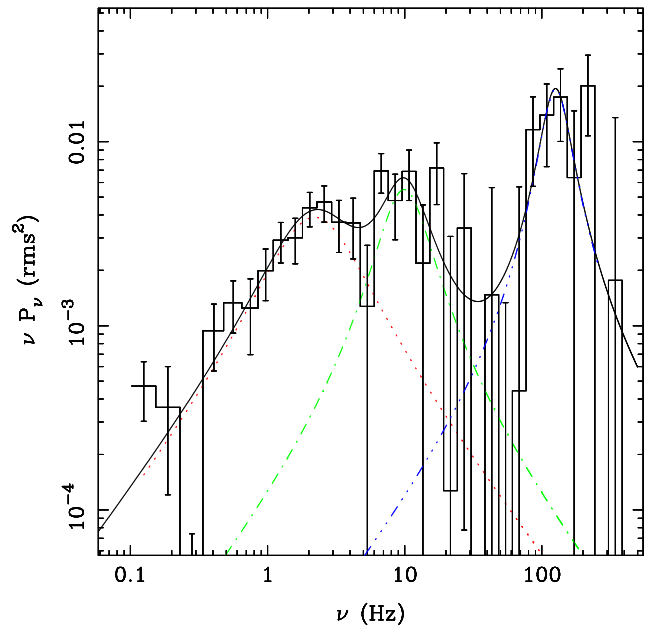}}}
\caption{Power density spectrum of XTE J1652$-$453 using the \textit{RXTE}/PCA data from the 
observation taken simultaneously with that of \textit{XMM-Newton} (MJD 55065, \textit{RXTE} 
ObsID 94432-01-04-00). The plot shows the frequency$\times$Fourier Power in units of squared 
rms fractional amplitude versus frequency. The black solid line is the best-fitting model to 
the data using three Lorentzians. The three Lorentzians are shown by the red dashed, green 
dot-dashed, and blue dot-dot-dashed lines, respectively.}
\label{fig:PDS}
\end{center}
\end{figure}
The power spectrum of the \textit{RXTE} observation that was taken simultaneously with the one of 
\textit{XMM-Newton},can be reasonably fit ($\chi^{2}=35.9$ for 25 d.o.f.) with three Lorentzians 
at $2.2 \pm 0.4$ Hz, $10.0 \pm 1.7$ Hz, and $131 \pm 20$ Hz, respectively. \citep[These are the 
frequencies at which the Lorentzians peak in the frequency$\times$power representation of e.g.,][]{Belloni02} 
The single-trial significance of each Lorentzian (defined as the power of the Lorentzian divided by 
its negative error) was 4.9$\sigma$, 2.8$\sigma$, and 3.4$\sigma$, respectively. The three Lorentzian 
had, respectively, quality factors, $Q=\nu_{0}$/FWHM (where $\nu_{0}$ and FWHM are the central 
frequency and the full-width at half maximum of the Lorentzian) of $0.5 \pm 0.2$, $1.0 \pm 0.7$, 
and $1.5 \pm 0.9$, and total fractional rms amplitudes of $8.9 \pm 0.9$, $8.3 \pm 1.6$, and $13.8 \pm 2.2$ 
per cent. In Fig.~\ref{fig:PDS} we show the power density spectrum of this observation, together with 
the best-fitting model. The black solid line is the total model, while the dashed, dotted and 
dashed-dotted lines in colour show the individual components.

If the marginally significant Lorentzian at $\sim 130$ Hz is real, we can use it to put an upper 
limit on the black-hole mass, as any frequency should be less than Keplerian. For a Keplerian 
frequency of $\nu \sim 130$ Hz at the inner edge of the accretion disc of $R_{\rm in} \sim 4$ $GM/c^2$, 
$\nu = c^{3}/[2\pi GM (R_{\rm in}^{3/2} + a)]$ \citep{Bardeen72}, with $a$ the spin parameter and 
$M$ the mass of the black hole, we find an upper limit on the mass of $\sim 30 M_{\odot}$. Given 
that most known BH binaries have mass around $\sim 10M_{\odot}$, then this shows that these 
fluctuations are indeed produced close to the black hole.
\section{Discussion}\label{sec:discussion}
The \textit{XMM-Newton}/\textit{RXTE} energy spectrum of XTE J1652$-$453 shows a broad and strong 
Fe emission line with an equivalent width of $\sim 450$ eV, which is among the strongest ever 
seen from a black-hole candidate \citep[see e.g.,][]{Rossi05,MillerChandra04}. The line is consistent with 
being produced by reflection off the accretion disc, broadened by relativistic effects. The power 
spectrum from \textit{RXTE} data can be fitted with two Lorentzians below 10 Hz, and shows an 
additional variability component above 50 Hz. The simultaneous \textit{XMM-Newton}/\textit{RXTE} 
observations were performed when XTE J1652$-$453 was in the hard-intermediate state \citep[HIMS;][]{Mendez97,Homan05}, 
as the outburst decayed from the disc dominated high/soft state towards the low/hard state. Our 
high-resolution, broadband observations substantially increase the number of good datasets from 
this state \citep[the other is GX 339$-$4]{MillerChandra04}. The strong line as well as the additional 
component at high frequencies are both characteristic of the hard-intermediate state. \citet{MillerChandra04} 
show a \textit{Chandra} grating spectrum of the black-hole candidate GX 339$-$4, with a very similar 
continuum spectrum as XTE J1652$-$453, where the line has an equivalent width of $\sim 600$ eV. 
\citet{KleinWolt08} show that there is an additional noise component at high frequencies in the 
HIMS, although it is somewhat weaker than the component seen in our data. Power spectra taken in 
the HIMS also typically show the strongest, most coherent, low-frequency QPO \citep[e.g.,][]{KleinWolt08}. 
The lack of any such coherent low-frequency QPO in our data may imply that we view the system at 
a rather low inclination angle. This is consistent with the inclination inferred from fitting the 
Fe line profile with relativistic models.%

Understanding the intermediate state is important as it marks the onset of the spectral transition, 
and the appearance of the radio jet. While there is still debate on the nature of the transition, 
most models currently explain it by the inner disc evaporating into a hot coronal flow \citep[see e.g.,][]{Liu06,Meyer00}. 
The intermediate state is therefore the one in which the disc is only slightly recessed from the 
last stable orbit, so there is strong disc emission, but there is also strong hard X-ray flux from 
the re-emerging hot flow. This geometry also predicts that there should be strong reflection features, 
and that these should be strongly smeared by relativistic effects. As the outburst declines further, 
the models predict that the inner disc quickly recedes, being replaced by the hot flow. With the 
receding disc, the solid angle subtended by the optically thick disc decreases, the amount of 
reflection decreases, and the larger radius means that the line is less strongly smeared. These 
models then suggest that the strongest and broadest Fe lines should be seen in the intermediate 
states, where the disc is closest to the last stable orbit, but co-exists with enough of the hard 
X-ray emitting corona to produce the required illumination of the disc.%

From low-resolution (\textit{RXTE}) spectral monitoring it is indeed seen that for the high/soft 
states, as there is very little hard X-ray emission illuminating the disc, the Fe lines are generally 
weak, while they appear strongly at the intermediate states, as coronal X-ray emission re-appear. 
The line width and strength then decrease as the outburst declines \citep{Gilfanov10}. However, 
such low-resolution data cannot distinguish changes in the reflection broadening from changes in 
the ionisation state, and since the disc temperature and/or strength of the irradiation change, 
the ionisation state should also change. Even if there is no change in geometry, an increase in 
disc temperature gives an increase in ionisation state, and so the strength of the consequent 
reflection features increases \citep{Ross93,Ross07}, as well as the line width increases due to 
increased Compton broadening \citep[e.g.,][]{Young05}. Moderate to high resolution data show that 
the Fe lines are ionised in the intermediate state \citep[e.g.,][and our data]{MillerChandra04}, and are 
predominantly neutral in the hardest low/hard state \citep[e.g.,][]{Tomsick09,Hiemstra09}.%

The high-resolution, broadband spectrum of XTE J1652$-$453 allows us to separately constrain ionisation,
solid angle and relativistic smearing. The ionisation is high with log$\xi \sim 3.5$, and the very broad 
Fe line requires strong relativistic effects. The fits using a self-consistent reflection model give 
$R_{\rm in} \simeq 4$ $GM/c^{2}$ corresponding to the last stable orbit for a black hole of spin $\sim 0.5$. 
This is a lower limit since we expect the disc to be slightly recessed. None the less, the \textit{RXTE} 
monitoring of XTE J1652$-$453 shows that the direct disc emission has similar temperature and normalisation 
as in the high/soft state earlier in the outburst \citepalias{ATel2107,ATel2120}, supporting the idea that the 
disc is very close to the last stable orbit in our data.%

However, the fits with the self-consistent reflection model show clear negative residuals at $\sim7.1$ keV. 
This could correspond to a blue-shifted absorption line, such as the ones that are often seen in high 
inclination systems \citep[e.g.,][]{Kubota07}. This absorption feature is not obviously apparent in 
the raw data, and it is only required when the smeared \textsc{reflionx} model is used since the 
intrinsic Compton broadening in this model predicts no sharp drop on the blue wing of the line 
\citep{Done06}. The \textsc{reflionx} model underestimates the Compton broadening as it calculates the
temperature of the disc from irradiation only, for $kT$ of 0.1--0.3 keV \citep{Ross93}. This is 
obviously lower than the observed inner disc temperature of 0.6 keV. The Compton broadened line width 
should be $\sigma \propto E_{\rm line}\sqrt{(2kT_{e})/(m_{e}c^2)}$ $\sim 0.32$ keV for He-like Fe.
This is negligible compared to the 1.7--2 keV broadening implied by $R_{\rm in} \sim 4$ $GM/c^{2}$, and 
should not significantly change the derived inner radius. The Compton broadening smooths out the sharp 
drop from the blue wing of the line even more, and increases the mismatch with the data at $\sim 7.1$ keV. 
Since the model does not describe the data in detail, we should be cautious in interpreting the reflection 
parameters in detail.
\section{Summary}\label{sec:summary}
We observed the black-hole candidate and X-ray transient XTE J1652$-$453 simultaneously with 
\textit{XMM-Newton} and \textit{RXTE}. During this observation the source was in the hard-intermediate 
state \citep{Homan05}. We found a strong and asymmetric Fe emission line in the X-ray spectrum 
of XTE J1652$-$453, with an equivalent width of $\sim 450$ eV, one of the strongest lines ever 
seen from a black-hole candidate. From fits to only \textit{XMM-Newton}/PN spectrum, the 
equivalent width is slightly lower, $\sim 380$ eV.

The line is also broad, giving an inner disc radius of $\sim 4$ $GM/c^{2}$, from either a single 
\textsc{Laor} line profile or from a self-consistent ionised reflection model. This implies a 
spin parameter of $\sim 0.5$, which is a lower limit as the disc may be slightly recessed 
in this state. In any case, neither model fits the data well in detail. This is not at all 
surprising for the single \textsc{Laor} line as the line must be accompanied by a reflected 
continuum. Be that as it may, the self-consistent ionised reflection model used here incorporates 
all the known effects such as multiple lines and edges from the different ionisation states 
of Fe (and other elements), together with Compton broadening \citep{Ross05}, such that the 
line is intrinsically broad. The characteristic skewed profile expected from a single narrow
line is smeared out, so there is no sharp drop in the model at the blue side of the line. 
However, the data clearly prefer such a feature, indicating that even the best current reflection 
models may not be able to describe the detailed shape of the Fe line seen in the spectrum of 
this state.

The Fourier power spectrum of the \textit{RXTE} observation simultaneous with that of 
\textit{XMM-Newton} shows additional variability above 50 Hz. No coherent QPOs at low frequency 
are accompanying this, which may suggest that we see the source at a fairly low inclination angle. 
The $3.4\sigma$ (single trial) component at $\sim 130$ Hz is too broad to be the high-frequency 
QPO, but in case it is, and assuming that it reflects the Keplerian frequency of the inner disc 
radius of $\sim 4$ gravitational radii (inferred from the line profile), then the estimated upper 
limit on the black-hole mass is $\sim 30$ $M_{\odot}$. None the less, since the source is in the 
hard-intermediate state rather than in the soft-intermediate state where the high-frequency QPOs 
are occasionally seen \citep{KleinWolt08}, this additional high-frequency component may be a 
related feature at lower frequencies and lower coherence, and it might be connected to the 
re-emergence of the jet.
\section{Acknowledgements}
We thank the \textit{XMM-Newton} team and the project scientist Norbert Schartel for performing the ToO 
observation. We also like to thank the \textit{RXTE} team for rescheduling to obtain a contemporaneous 
observation. We are grateful to an anonymous referree for constructive comments, and we thank Juri Poutanen 
for a valuable discussion. PC acknowledges funding via a EU Marie Curie Intra-European Fellowship under 
contract no. 2009-237722
\bibliographystyle{mn}
\bibliography{./biblio}
\appendix
\section{Choice of background and abundances}
\subsection{Background in PN timing mode}\label{app:background}
As can be seen in Fig.~\ref{fig:src-vs-bkg}, the PN background spectrum (red) extracted 
from the XTE J1655$-$453 observation, is contaminated with source photons (see 
Section~\ref{sec:XMM-Newton} for more details about the extraction criteria). Using this 
background spectrum leads to an over-subtraction of the source spectrum and it may change 
the results of the spectral analysis. To investigate whether the continuum and line parameters 
are affected by the background correction, we performed three fits to the source spectrum of 
XTE J1652$-$453, using a different background correction in each fit. 
\setcounter{figure}{6}
\begin{figure}
\begin{center}
\resizebox{\columnwidth}{!}{\rotatebox{-90}{\includegraphics[clip]{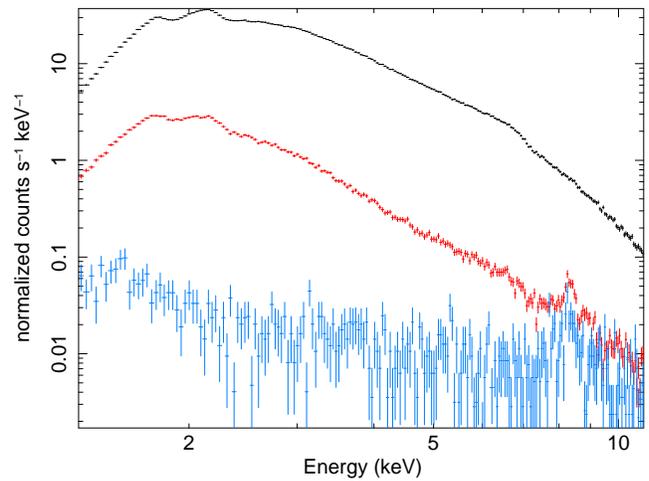}}}
\caption{Source and background spectra extracted from \textit{XMM-Newton} data with PN in 
timing mode. The background and source spectra are all rebinned using the tool \textsc{pharbn} 
(see Section~\ref{sec:XMM-Newton}). The \textit{black} (top) spectrum is the source spectrum 
of XTE J1652$-$453, extracted from the RAWX columns 30 to 46; the \textit{red} (middle) 
spectrum is the background spectrum of XTE J1652$-$453 using RAWX in [5:21]; and the \textit{blue} 
(bottom) spectrum is the background spectrum extracted from a `blank-sky field' observation 
of 4U 1608$-$52 using the RAWX columns 10 to 26.}
\label{fig:src-vs-bkg}
\end{center}
\end{figure}
As a first background correction (Model 1a) we used the background spectrum extracted from 
the outer parts of the CCD of the same observation of XTE J1652$-$453 (RAWX in [5:21]). This
is the background spectrum contaminated with source photons. For the second background 
correction (Model 1b) we extracted a spectrum from a so-called blank-sky field observation 
that was taken with PN in timing mode, and as our last correction (Model 1c) we did not 
subtract any background.

Since the background can be different in different parts of the sky, for a blank-sky field 
we chose an observation taken in a direction close to that of XTE J1652$-$453; we therefore
used the observation of the neutron-star low-mass X-ray binary 4U 1608$-$52 (obsID 0074140201) 
taken just after an outburst, where the source had returned to quiescence. From all the 
blank-sky field observations taken in a direction close to XTE J1652$-$453, the one of 4U 
1608$-$52 has the most similar $N_{\rm H}$ value compared with XTE J1652$-$453. Since in 
the observation of 4U 1608$-$52 the source was still detectable, although it was weak, we 
extracted the background spectrum using the columns RAWX in [10:26], confirming that we do 
not include any columns contaminated with source photons. In Fig.~\ref{fig:src-vs-bkg} we 
show the background spectrum extracted from the 4U 1608$-$52 observation in blue. As can be 
seen from both background spectra in Fig.~\ref{fig:src-vs-bkg}, there are emission features 
at $\sim 8$ keV that are due to fluorescence from the detectors and the structure surrounding 
them. The lines at $\sim 8$ keV are the fluorescence lines of Cu-K$\alpha$ and Zn-K$\alpha$ 
(see the \textit{XMM-Newton} Users Handbook), and are time-dependent. Using a background 
spectrum from a different observation than the source spectrum, as we do for 4U 1608$-$52, 
may cause an incorrect background subtraction at around $\sim 7$--8 keV and may change the 
parameters of a possible Fe line at $\sim 7$ keV.

For the three spectral fits with different background corrections, we fit Model 1 to PN data 
only, covering the 1.3--10 keV range, and excluding the 1.75--2.25 keV range. As described in 
Section~\ref{sec:spectroscopy}, we added an \textsc{edge} at $\sim 9.2$ keV. The results for 
the different background corrections are given in Table~\ref{tab:backgrounds}.

We set the abundances according to values reported in \citet{Anders89}, and we used cross 
sections of \citet{Verner96}. For the purpose of comparing different background corrections, 
the assumed abundances and cross sections are not important. However, the best-fitting 
parameters and their physical interpretation can be affected when different abundance/cross-section 
tables are used (see following Section). For comparison, in Table~\ref{tab:backgrounds} we 
also give the best-fitting results of Model 1c, where we instead used the abundances of 
\citet{Wilms00} (Model 1d).
\setcounter{table}{1}
\begin{table*}
\begin{minipage}{140mm}
\begin{center}{
\scriptsize \caption{Spectral fit results for different background corrections}\label{tab:backgrounds}
\begin{tabular}{clcccc}

\multicolumn{1}{c}{Model component} & 
\multicolumn{1}{c}{Parameter} & 
\multicolumn{1}{c}{\hrulefill~Model 1a~\hrulefill} &  
\multicolumn{1}{c}{\hrulefill~Model 1b~\hrulefill} &  
\multicolumn{1}{c}{\hrulefill~Model 1c~\hrulefill} &
\multicolumn{1}{c}{\hrulefill~Model 1d~\hrulefill}\\\hline\hline

\textsc{phabs}   & $N_{\rm H}$ ($10^{22}$)     & 4.75$\pm$0.03          & 4.66$\pm$0.04          & 4.64$\pm$0.03          & 6.78$\pm$0.03      \\ \hline

\textsc{diskbb}  & $kT_{\rm in}$ (keV)         & 0.56$\pm$0.01          & 0.56$\pm$0.01          & 0.56$\pm$0.01          & 0.58$\pm$0.01      \\
                 & $N_{\rm dbb}$               & 1299$^{+95}_{-33}$     & 1428$^{+95}_{-81}$     & 1390$^{+52}_{-84}$     & 1060$^{+24}_{-45}$ \\\hline

\textsc{powerlaw}& $\Gamma$                    & 2.52$\pm$0.09          & 2.52$\pm$0.09          & 2.47$\pm$0.05          & 2.34$\pm$0.07      \\
                 & $N_{\rm \Gamma}$            & 0.33$\pm$0.06          & 0.33$\pm$0.06          & 0.30$\pm$0.06          & 0.23$\pm$0.04      \\\hline

\textsc{Laor}    & $E_{\rm L}$ (keV)           & 6.97$^{+0}_{-0.02}$    & 6.97$^{+0}_{-0.03}$    & 6.97$^{+0}_{-0.04}$    & 6.97$^{+0}_{-0.02}$   \\
                 & $\beta$                     & 3.3$\pm$0.1            & 3.3$\pm$0.1            & 3.3$\pm$0.1            & 3.4$\pm$0.1           \\
                 & $R_{\rm in}$ ($GM/c^{2}$)   & 4.01$^{+0.25}_{-0.05}$ & 4.01$^{+0.29}_{-0.46}$ & 4.01$^{+0.23}_{-0.13}$ & 4.01$^{+0.19}_{-0.07}$\\
                 & $\theta$ (deg)              & 6.1$^{+3.4}_{-5.0}$    & 4.4$^{+6.3}_{-4.4}$    & 3.5$^{+7.0}_{-3.5}$    & 5.1$^{+2.9}_{-5.1}$   \\
                 & $N_{\rm L}$ ($10^{-3}$)     & 1.2$\pm$0.1            & 1.2$\pm$0.3            & 1.2$\pm$0.1            & 1.3$\pm$0.1           \\
                 & EW (eV)                     & 352$^{+39}_{-24}$      & 348$^{+73}_{-32}$      & 343$^{+36}_{-29}$      & 375$^{+29}_{-24}$     \\ \hline

           & $\chi^{2}_{\nu}$ ($\chi^{2}/\nu$) & 1.58 (261.0/165)       & 1.49 (246.6/165)       & 1.45 (240.0/165)       & 1.50 (247.4/165)\\
           & Total flux ($10^{-9}$)            & 8.99                   & 9.31                   & 9.28                   & 8.87            \\\hline\hline

\multicolumn{6}{l}{~~~NOTES.--~Results are based on spectral fits to PN data of XTE J1652$-$453 in the 1.3--10 keV range. Background cor-}\\
\multicolumn{6}{l}{rections are: background of XTE J1652$-$453 contaminated with source photons (Model 1a); background of 4U 1608$-$52}\\ 
\multicolumn{6}{l}{(Model 1b), and no background subtracted (Model 1c \& 1d). ~The flux is unabsorbed and given for the 2--10 keV range,}\\
\multicolumn{6}{l}{in units of erg cm$^{-2}$ s$^{-1}$. Cross sections are set to \citet{Verner96}, and abundances to \citet{Anders89}}\\
\multicolumn{6}{l}{for Models 1a--c, and to \citet{Wilms00} for Model 1d. Details about parameters are given in Section~\ref{sec:spectroscopy}.}
\end{tabular}
\normalsize}
\end{center}
\end{minipage}
\end{table*}

We find that for Models 1a--c \citep[i.e., with abundances of][]{Anders89} all continuum parameters, 
except for $N_{\rm H}$, are consistent between the three different background corrections. The column 
density in Model 1a is slightly higher than in Models 1b and 1c. The \textsc{diskbb} normalisation is
slightly lower in Model 1a than in Models 1b and 1c, but is consistent within errors. The higher 
$N_{\rm H}$ can be due to the over-subtraction of the background, which also explains the lower $N_{\rm dbb}$ 
\citep[see also][for discussion about $N_{\rm H}$ versus $N_{\rm dbb}$ relation]{Cabanac09}. 
We conclude that for the case of XTE J1652$-$453 the choice of background correction has no effect
on the continuum and line parameters. However, the issue of a contaminated background would be 
relevant if the \textsc{diskbb} normalisation is used to determine the inner disc radius. 

The $\chi^{2}_{\nu}$ given in Table~\ref{tab:backgrounds} are relatively high, although the parameters 
obtained from fits to only PN data are consistent with the fits to \textit{XMM-Newton}/\textit{RXTE} 
data, where the reduced $\chi^{2}$ is acceptable. The contribution to the high $\chi^{2}_{\nu}$ is 
most likely due to uncertainties in the detector calibration. An additional systematic error of 0.6 
per cent to all PN channels is sufficient to get a $\chi^{2}_{\nu} \sim 1$.
\subsection{Abundances, cross sections, and energy coverage}\label{app:nh}
By investigating the effects of the different background corrections, we have used different abundances 
than in the rest of our analysis. By comparing the best-fitting results of Model 1d with those of 
Models 1a--c in Table~\ref{tab:backgrounds}, we find that the equivalent hydrogen column density is 
$\sim 30$ per cent lower if we used the abundances of \citet{Anders89} than if we used the abundances of 
\citet{Wilms00}. With the change in $N_{\rm H}$, the \textsc{diskbb} normalisation is significantly 
affected, and with this the other continuum parameters are slightly changed. Further, the best-fitting 
parameters of Model 1d, although slightly different, are consistent with those for Model 1 
(Table~\ref{tab:line-models}). In other words, fitting the continuum spectrum in a relatively narrow 
band (1.3--10 keV, only PN) or in a wide-coverage (0.4--150 keV) did not change the parameters 
significantly. However, fits to only PN data resulted in larger errors on the \textsc{powerlaw} 
parameters, and also resulted in a higher \textsc{powerlaw} normalisation (although still consistent 
within errors) than for the broadband fits. With the higher \textsc{powerlaw} normalisation the
equivalent width of the Fe emission line is clearly lower in the fits to PN data only than for the 
broadband fits. Nevertheless, for the case of XTE J1652$-$453 we found that even for the different 
continuum parameters, the Fe line parameters are not affected significantly. The best-fitting line 
parameters are consistent for the different background corrections, abundances, and energy coverage 
that we used.

\bsp
\label{lastpage}

\end{document}